\documentclass[aip,jcp,amsmath,amssymb,reprint]{revtex4-1}
\usepackage{graphicx}
\usepackage{amssymb}
\usepackage{amsmath}
\usepackage{color}
\usepackage[bookmarks=false]{hyperref}
\usepackage{bm}
\usepackage{upgreek}
\usepackage[multidot]{grffile}
\usepackage{tikz}
\usepackage{relsize}
\usepackage{caption}
\usepackage{subcaption}
\usepackage[utf8]{inputenc}
\usepackage{changes}
\usepackage{tabularx}

\tikzset{fontscale/.style = {font=\relsize{#1}}}

\begin{document}

\title{Assembling Ellipsoidal Particles at Fluid Interfaces using Switchable Dipolar Capillary Interactions}

\author{Gary B. Davies}
\email{g.davies.11@ucl.ac.uk}
\affiliation{Centre for Computational Science, University College London, 20 Gordon Street, London WC1H 0AJ, United Kingdom.}

\author{Timm Kr\"uger}
\email{timm.krueger@ed.ac.uk}
\affiliation{Institute for Materials and Processes, School of Engineering, University of Edinburgh, Mayfield Road, Edinburgh EH9 3JL, Scotland, United Kingdom.}
\affiliation{Centre for Computational Science, University College London, 20 Gordon Street, London WC1H 0AJ, United Kingdom.}

\author{Peter Coveney}
\email{p.v.coveney@ucl.ac.uk}
\affiliation{Centre for Computational Science, University College London, 20 Gordon Street, London WC1H 0AJ, United Kingdom.}

\author{Jens Harting}
\email{j.harting@tue.nl}
\affiliation{Department of Applied Physics, Eindhoven University of Technology, P.O. Box 513, 5600
MB Eindhoven, The Netherlands.}
\affiliation{Faculty of Science and Technology, Mesa+ Institute, University of
Twente, 7500 AE Enschede, The Netherlands.}

\author{Fernando Bresme}
\email{f.bresme@imperial.ac.uk}
\affiliation{Department of Chemistry, Imperial College London, London, SW7 2AZ, United Kingdom.}
\affiliation{Department of Chemistry, Norwegian University of Science and Technology, Trondheim, Norway.}

\pacs{}
\maketitle

The fabrication of novel soft materials is an important scientific and technological challenge. We investigate the response of ellipsoidal particles adsorbed at fluid-fluid interfaces to external magnetic fields. By exploiting previously discovered first-order orientation phase transitions,~\cite{davies_effect_2014,bresme_orientational_2007,bresme_computer_2008} we show how to switch on and off dipolar capillary interactions between particles, leading to the formation of distinctive self-assembled structures and allowing dynamical control of the bottom-up fabrication of novel-structured materials.\\\indent
Particles adsorb strongly at fluid-fluid interfaces: detachment energies of spherical particles can be orders of magnitude greater than the thermal energy, $k_B T$.~\cite{binks_particles_2002,bresme_nanoparticles_2007}
Once particles are adsorbed at an interface, competing hydrodynamic, electromagnetic and capillary interactions can lead to particles self-assembling into materials with specific mechanical, optical, or magnetic properties.~\cite{Madivala2009b,ozin2009nanochemistry}\\\indent
Capillary interactions~\cite{botto_capillary_2012} have attracted interest for their role in assembling mosquito eggs adsorbed at air-water interfaces,~\cite{loudet_how_2011} suppressing the coffee ring effect,~\cite{yunker_suppression_2011} and aggregating Cheerios.~\cite{larmour_sheets_2008,vella_cheerios_2005} Capillary interactions occur because of the creation of menisci around particles, and these particle-induced interface deformations, called capillary charges, interact with one another analogously to electric charges.~\cite{kralchevsky_capillary_1994, kralchevsky_capillary_2000}

The height of the interface deformation, $h$, due to the presence of a particle obeys Young's equation, $\nabla^2 h = 0$, which can be solved using a multipole expansion.~\cite{botto_capillary_2012,lehle_ellipsoidal_2008,oettel_colloidal_2008} Heavy particles deform the interface symmetrically in a monopolar fashion leading to $\log r$ interaction potentials between particles, where $r$ is the inter-particle distance.~\cite{vella_cheerios_2005,larmour_sheets_2008,botto_capillary_2012} However, for micron-sized particles, gravitational forces are usually small compared to surface-tension forces.~\cite{botto_capillary_2012} 

Non neutrally-wetting micron-sized ellipsoidal particles adsorbed at fluid-fluid interfaces contort the interface around them in a quadrupolar fashion purely because of their shape:~\cite{loudet_capillary_2005, botto_capillary_2012} the interface is depressed more at the tips than it is elevated at the sides, leading to orientation-dependent interaction potentials between particles.~\cite{lehle_ellipsoidal_2008,botto_capillary_2012,loudet_capillary_2005,oettel_colloidal_2008} The resulting capillary interaction energies can be several orders of magnitude larger than the thermal energy, $E \sim 10^5 k_B T$, providing a strong driving force for self-assembly.~\cite{loudet_capillary_2005, botto_capillary_2012} However, neither the monopolar nor the quadrupolar interactions are dynamically tunable.

\begin{figure}[b]

	\includegraphics{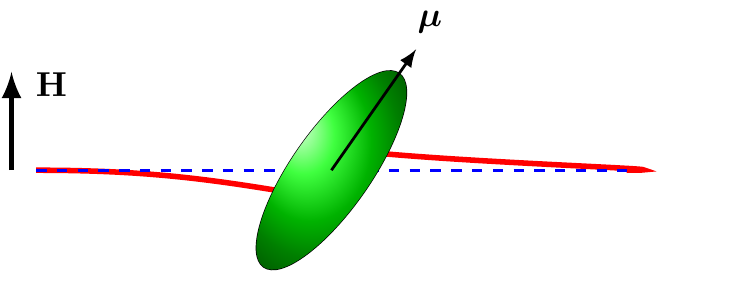}
	\\ [-0.3cm]
	\caption{\raggedright Anti-symmetric meniscus formation (solid red lines) for a single magnetic ellipsoidal particle with dipole moment, $\boldsymbol{\mu}$, under the influence of an applied magnetic field, $\mathbf{H}$, directed normal to the undeformed interface (blue dashed).}
	\label{img:single_particle}
	
\end{figure}

\begin{figure*}
	\textbf{Before First-Order Transition \hspace{0.83cm} 	\begin{tikzpicture}
               \node at (0.0,0.0) [label=above:{$\boldsymbol{H}$}] {};
		        \draw [black, ultra thick] (0.0,0.0) circle [radius=0.15];
			\draw [black, ultra thick,fill] (0.0,0.0) circle [radius=0.02];
        \end{tikzpicture}  
  \hspace{1cm} After First-Order Transition}

        \centering
        \begin{subfigure}[t]{0.22\textwidth}
                \includegraphics[width=0.96\textwidth]{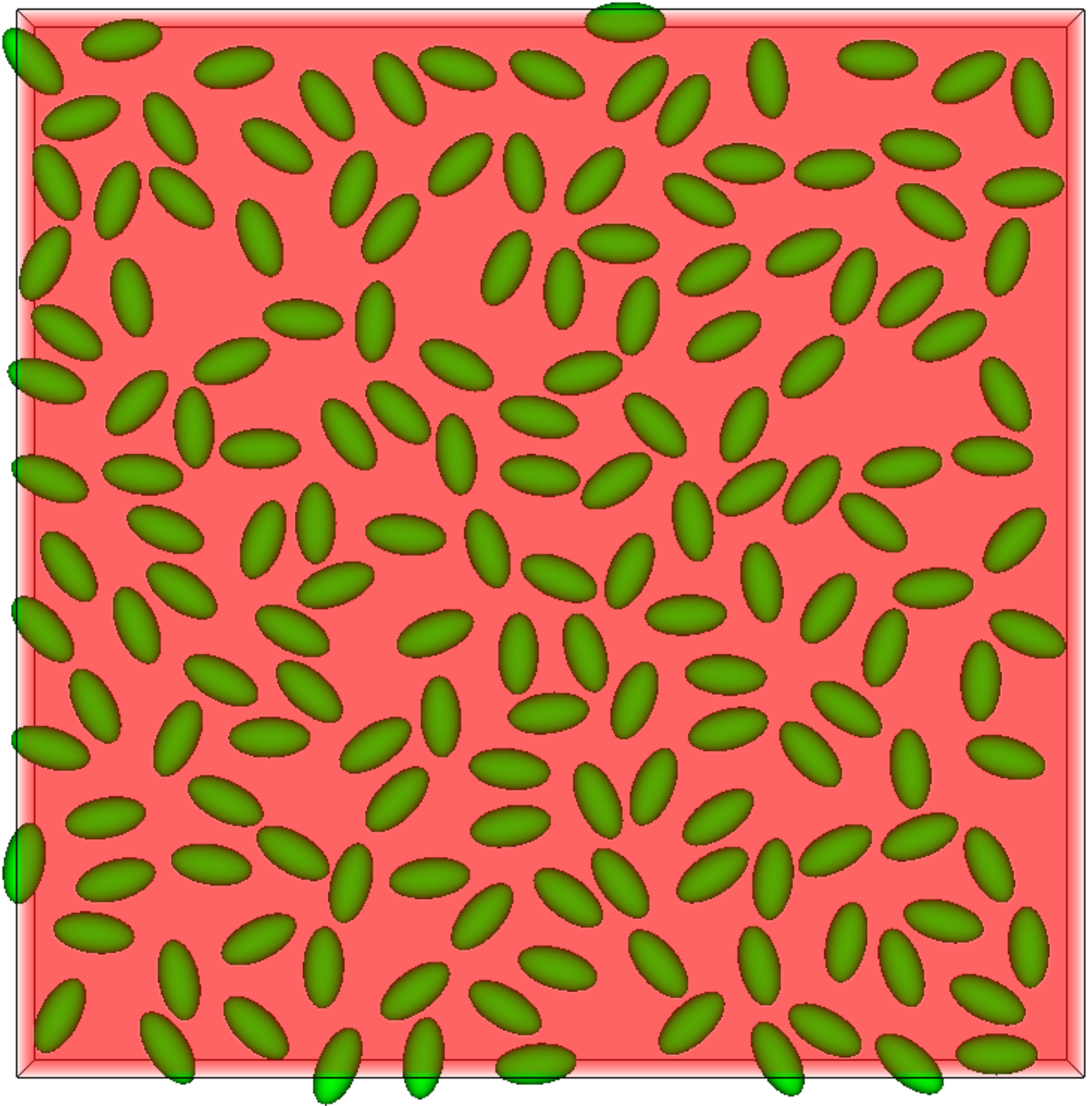}
		\caption{$B = 0$, $t = 0$}
                \label{img:init}
        \end{subfigure}
        ~ 
        \begin{subfigure}[t]{0.22\textwidth}
                \includegraphics[width=\textwidth]{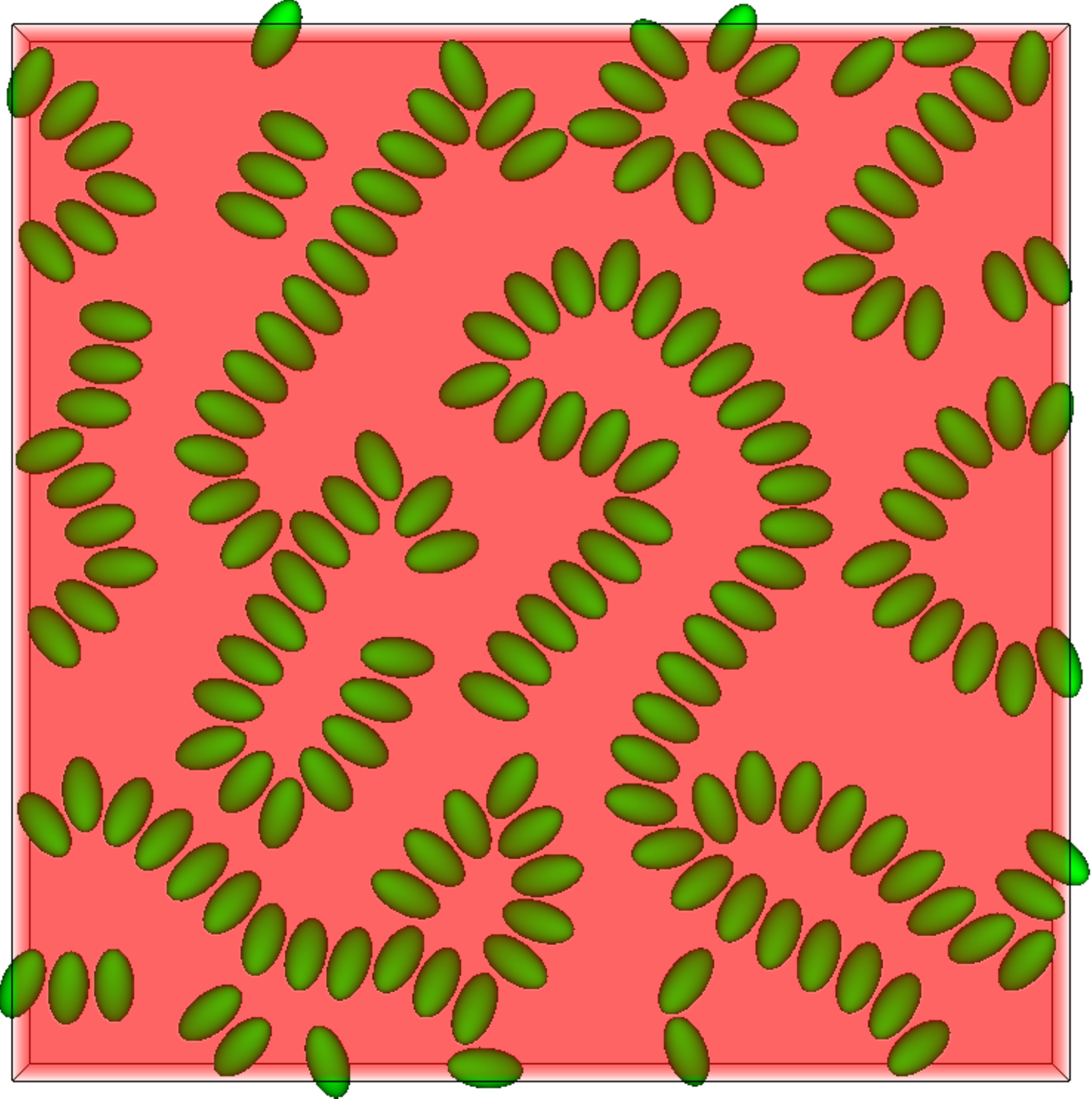}
                \caption{$B=\frac{1}{2}B_c$, $t \rightarrow \infty$}
                \label{img:ordered1}
        \end{subfigure}
        ~
        \begin{subfigure}[t]{0.000001\textwidth}
        \begin{tikzpicture}
        \draw[dashed,thick] (0,0) -- (0,3.9);
        \end{tikzpicture}
        \end{subfigure}
        ~ 
        \begin{subfigure}[t]{0.22\textwidth}
               \includegraphics[width=0.985\textwidth]{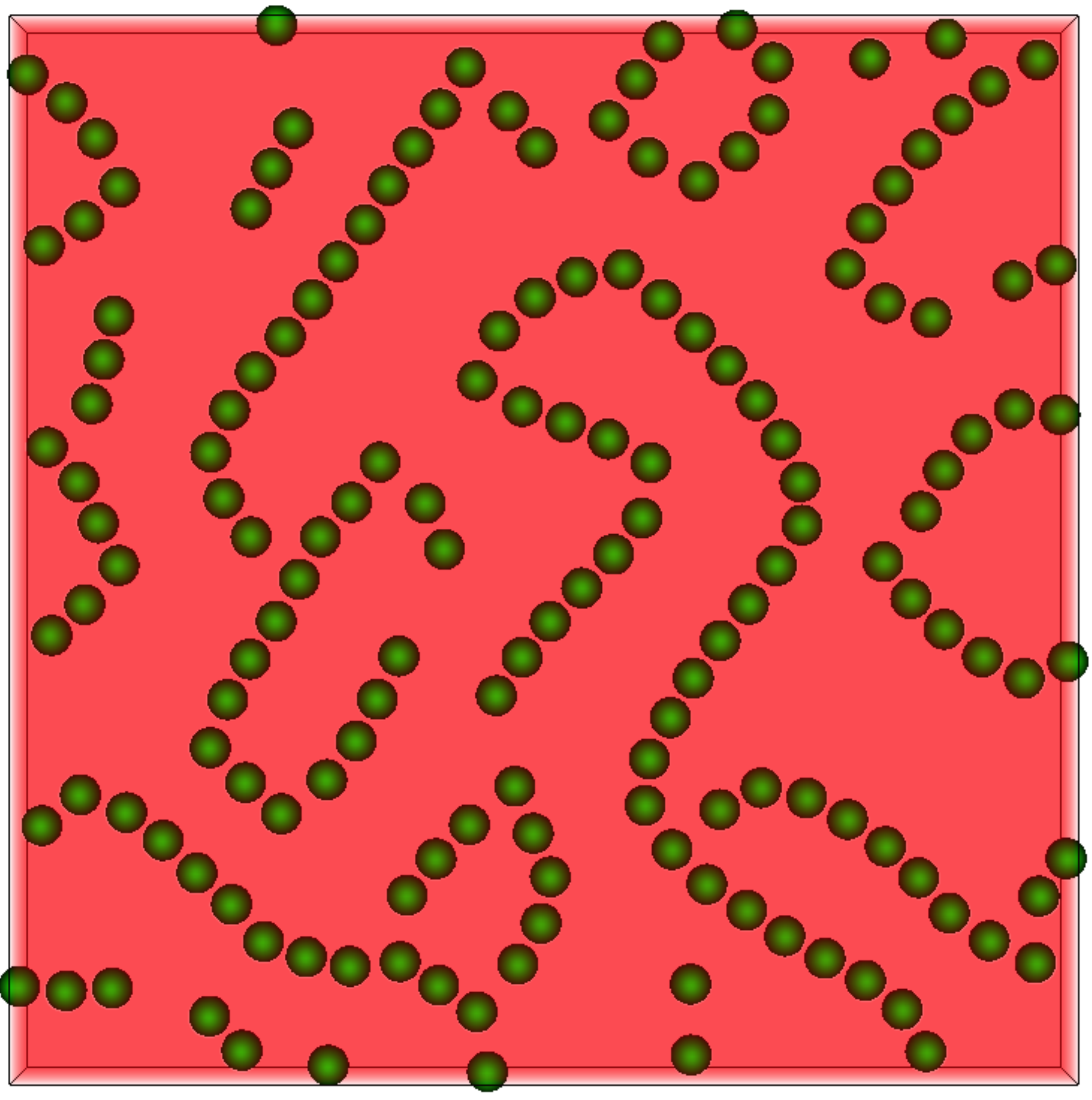}
                \caption{$B > B_c$, $t = 0$}
                \label{img:ordered2}
        \end{subfigure}
         ~ 
        \begin{subfigure}[t]{0.22\textwidth}
                \includegraphics[width=0.982\textwidth]{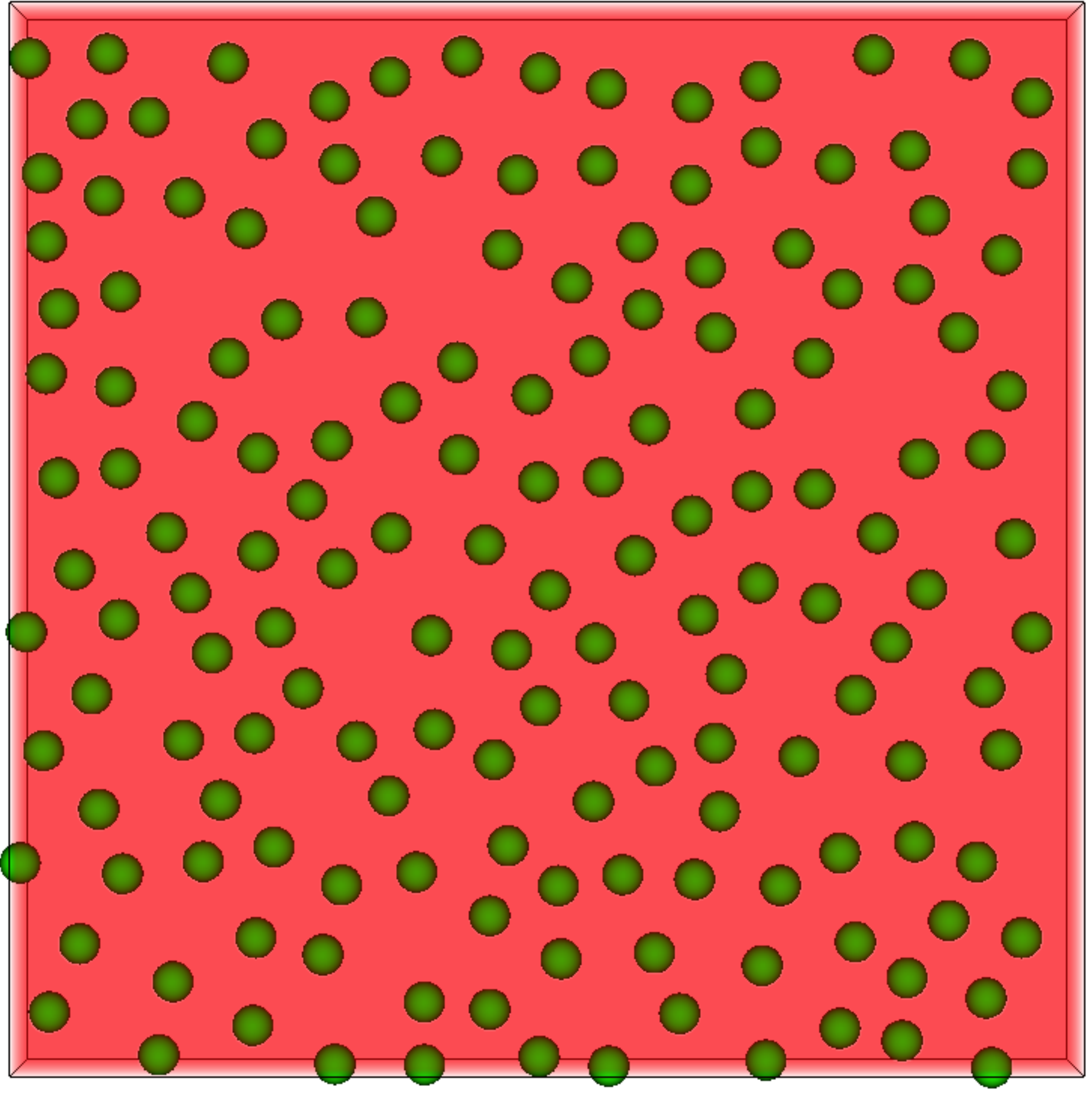}
                \caption{$B > B_c$, $t \rightarrow \infty$}
                \label{img:disordered}
        \end{subfigure}
        \\ [-0.3cm]
        \caption{\raggedright (a) The particles are distributed randomly in their equilibrium orientations with surface fraction $\phi=0.38$. (b) Applying a magnetic field parallel to the interface normal, $\boldsymbol{H}$, causes them to self-assemble due to dipolar capillary interactions. (c) Once the critical field strength is reached, particles transition to the vertical state, halting dipolar capillary interactions. (d) Illustration: once capillary interactions have been turned off, the particles may order randomly if magnetic dipole-dipole and van der Waals interactions are weak compared to thermal fluctuations.}
        \label{fig:self_assembly}
\end{figure*}
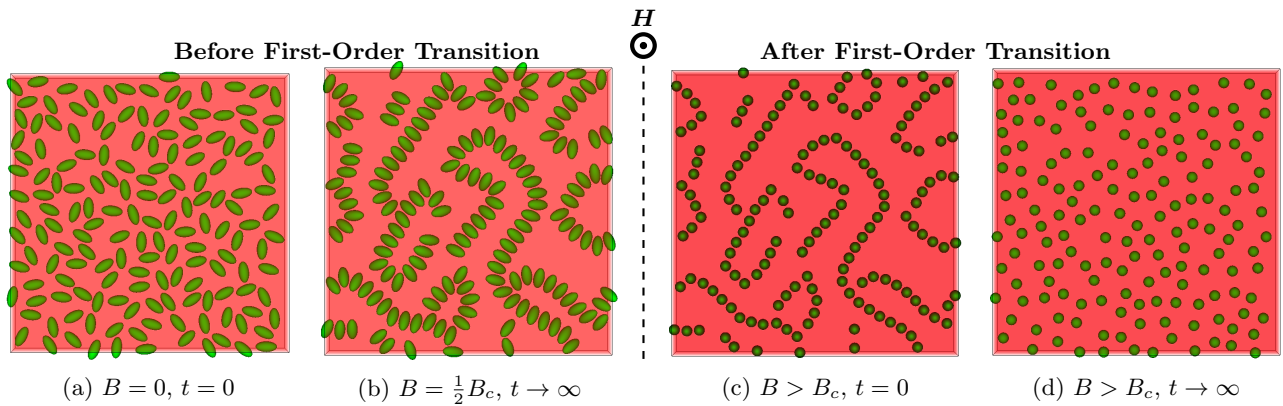

Dipolar capillary charges are created by anisotropic particles influenced by torques. External torques can be caused by particles with uneven mass distributions interacting with gravity, complex surface chemistries (e.g. Janus particles) interacting with fluids,~\cite{li_size-controlled_2013,guell_magnetically_2011,liu_self_2012} or --- the focus of this Communication --- embedded dipoles interacting with external magnetic fields, opening up new routes for the manipulation of particle monolayers. 

Materials science advances~\cite{snoeks_colloidal_2000,Madivala2009b} have enabled the production of anisotropic particles with embedded ferromagnetic dipoles~\cite{zabow_ellipsoidal_2014} or (super)-paramagnetic dipoles~\cite{hyeon_chemical_2003,li_colloidal_2011} so that particles can interact with external magnetic fields.

Bresme et al.~\cite{bresme_orientational_2007} investigated the behaviour of magnetic ellipsoidal particles adsorbed at fluid-fluid interfaces under the action of a magnetic field, predicting that particles undergo a discontinuous, first-order phase transition from a tilted state to a vertical state if a critical dipole-field strength is reached.~\cite{bresme_computer_2008} Davies et al.~\cite{davies_effect_2014} provided further evidence that the transition exists and also showed that particle-induced interface deformations significantly affect the transition.

\begin{figure*}

\begin{center}
\begin{tabular}{m{0.15cm}m{3.4cm}m{3.4cm}m{3.4cm}m{3.4cm}m{3.4cm}}

& \begin{center} 0.0$B_c$ \end{center} & \begin{center} 0.2$B_c$ \end{center} & \begin{center} 0.5$B_c$ \end{center} & \begin{center} 0.8$B_c$ \end{center} & \begin{center} 1.2$B_c$ \end{center} \\ [-0.4cm]

\rotatebox{90}{\hspace{1.5cm}$\phi=0.38$}
&\includegraphics[scale=0.135]{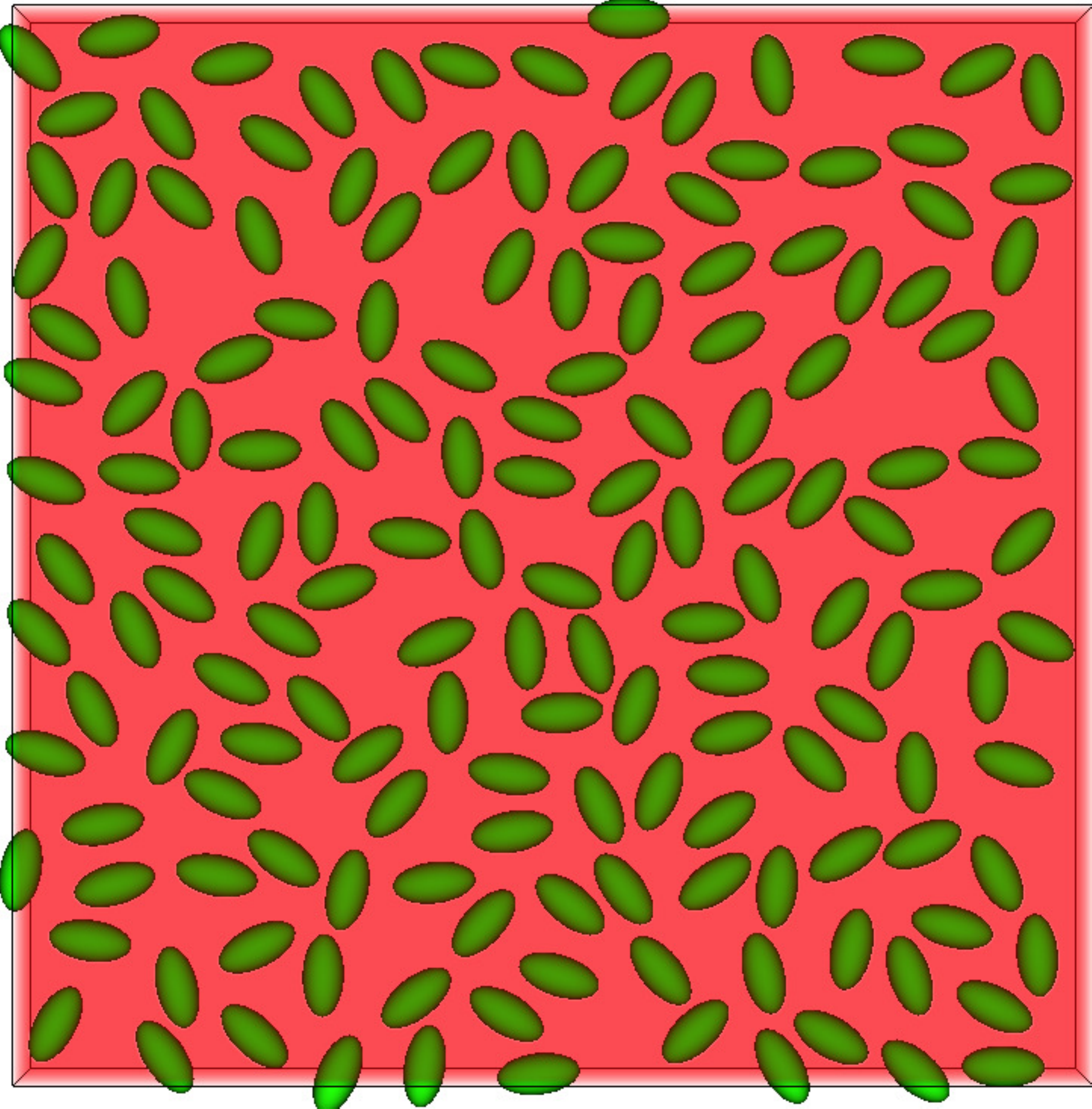}
\subcaption{} 
&\includegraphics[scale=0.135]{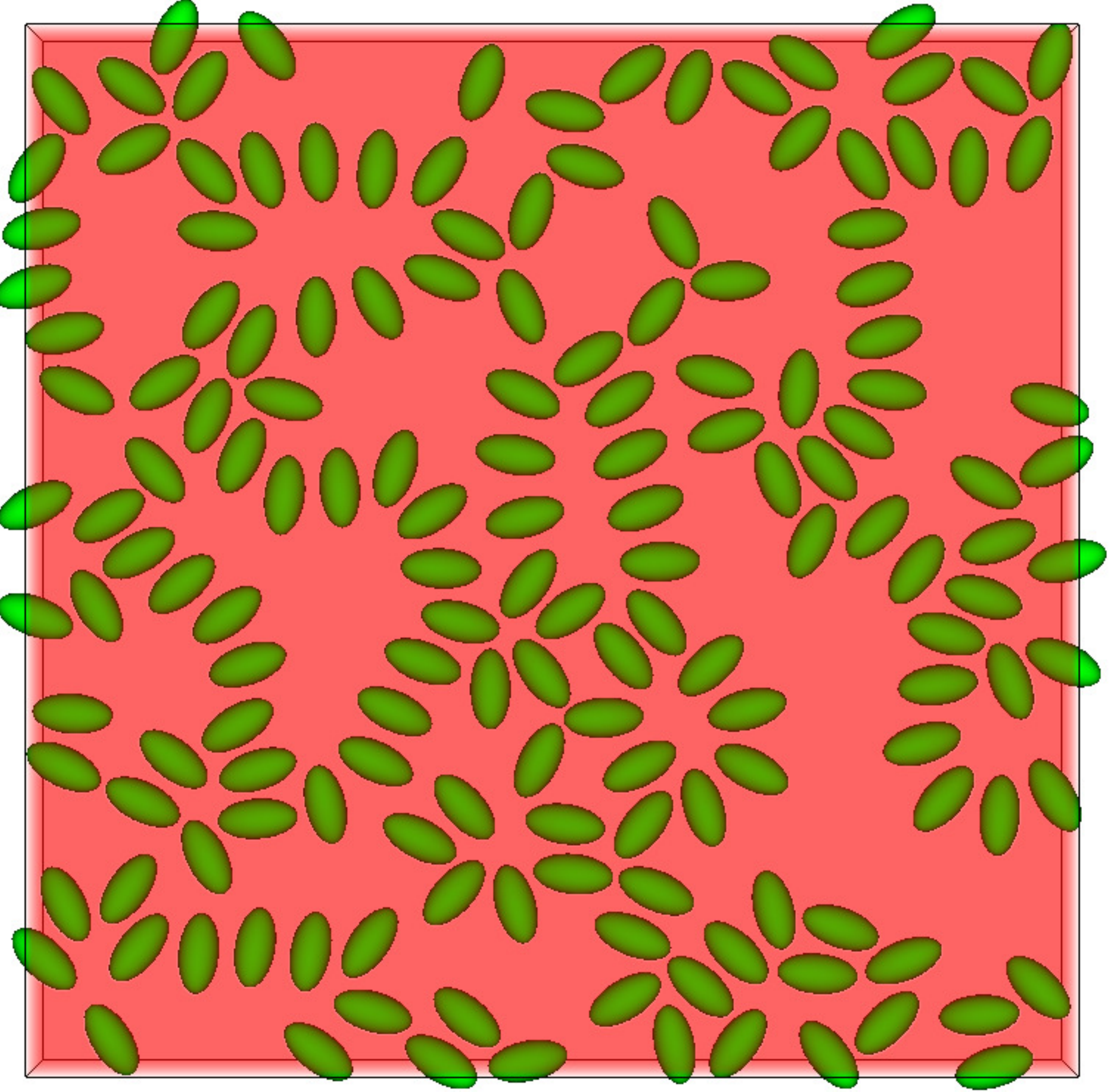}
\subcaption{} 
&\includegraphics[scale=0.135]{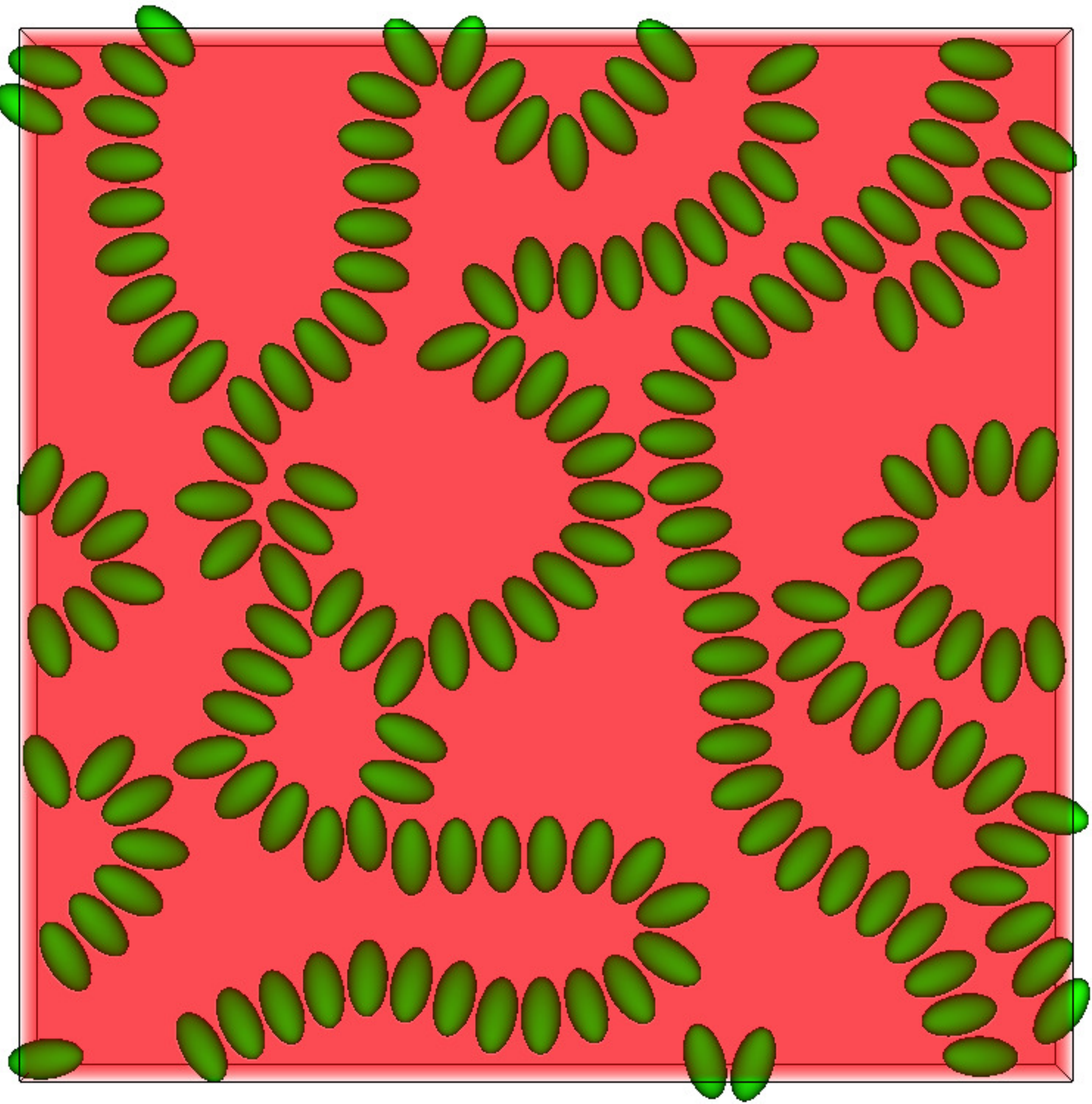} 
\subcaption{} 
&\includegraphics[scale=0.135]{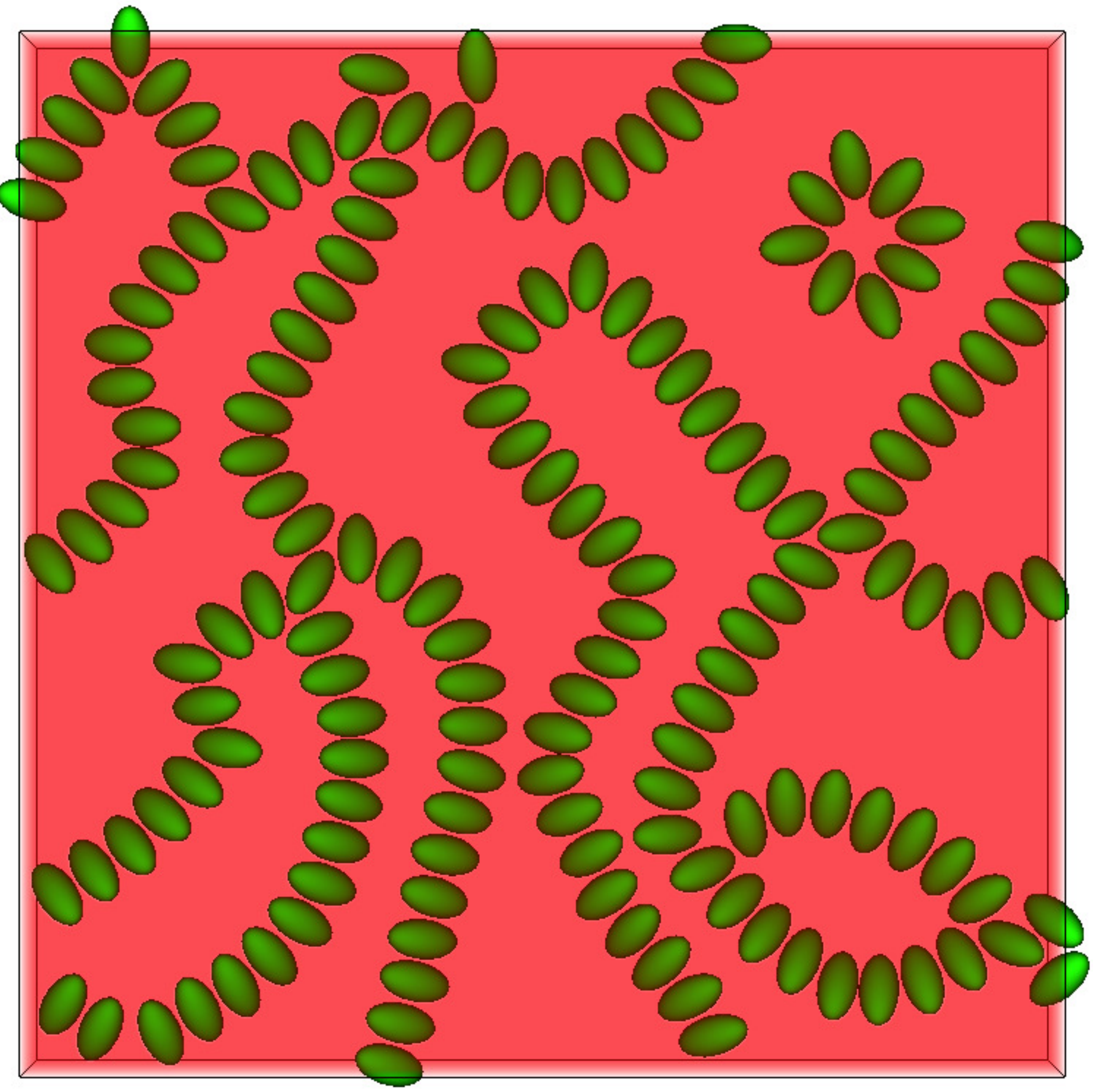} 
\subcaption{} 
&\includegraphics[scale=0.135]{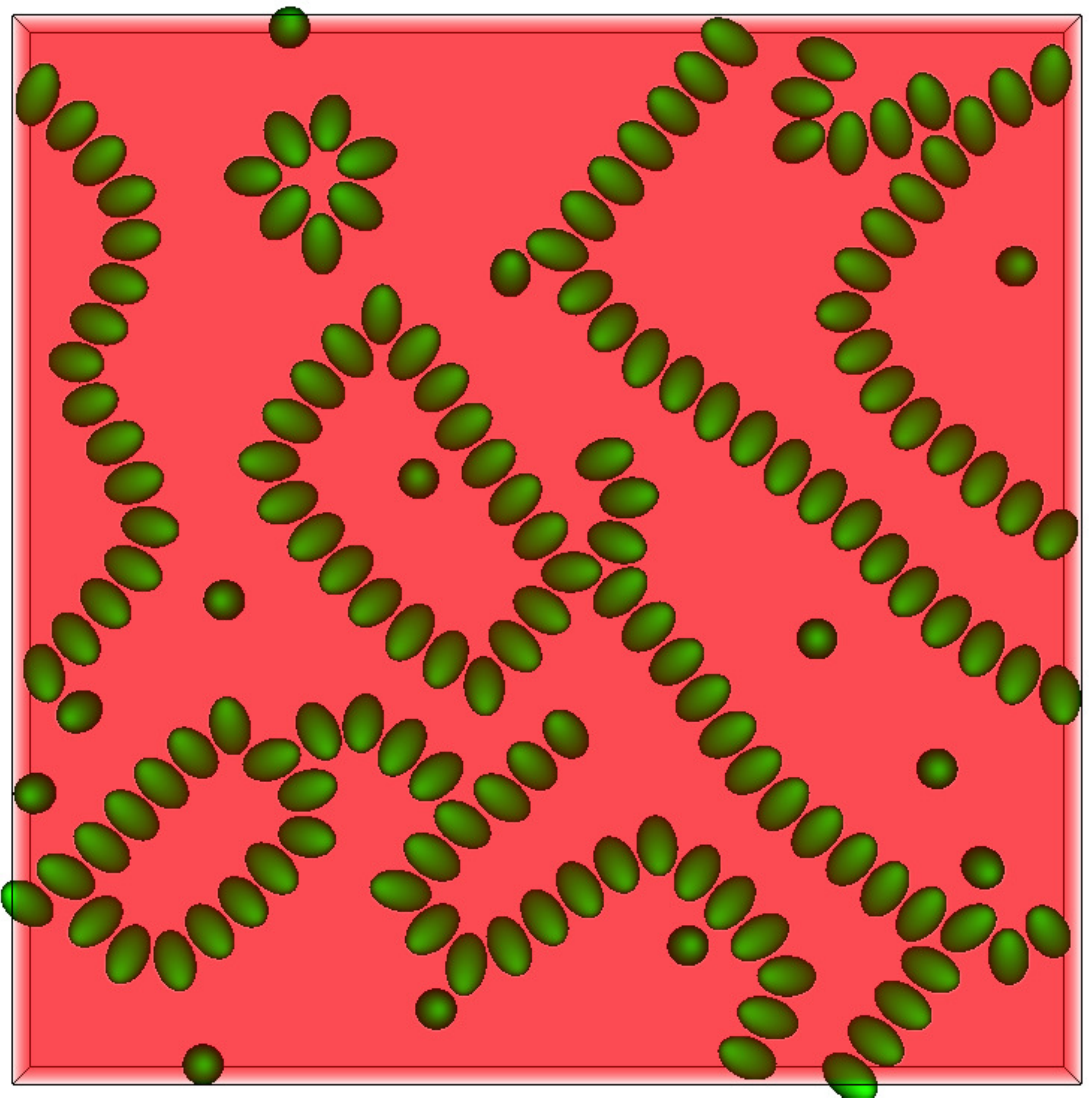}
\subcaption{} \\ [-0.9cm]
\rotatebox{90}{\hspace{1.5cm}$\phi=0.53$}
&\includegraphics[scale=0.135]{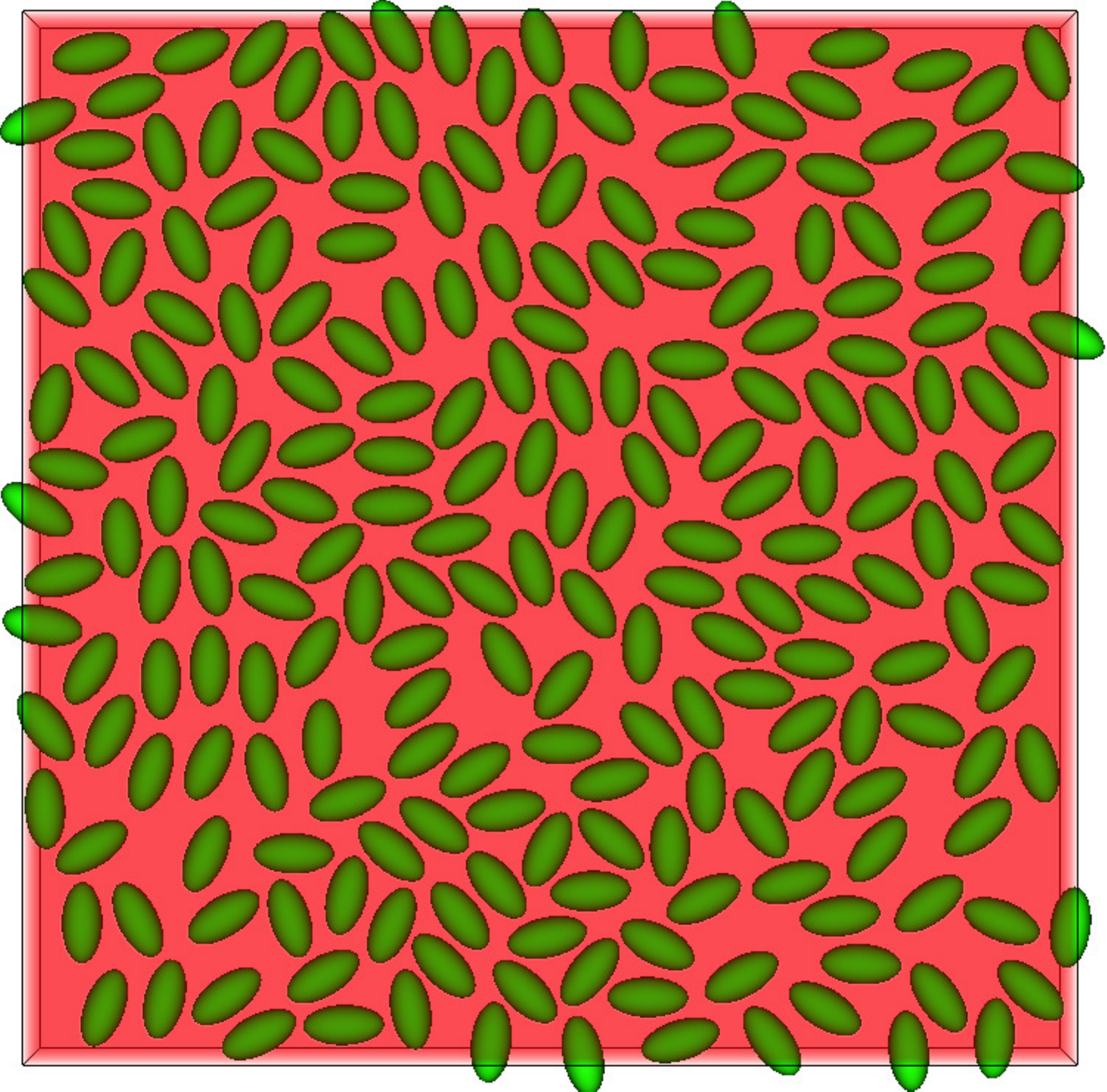}
\subcaption{} 
&\includegraphics[scale=0.135]{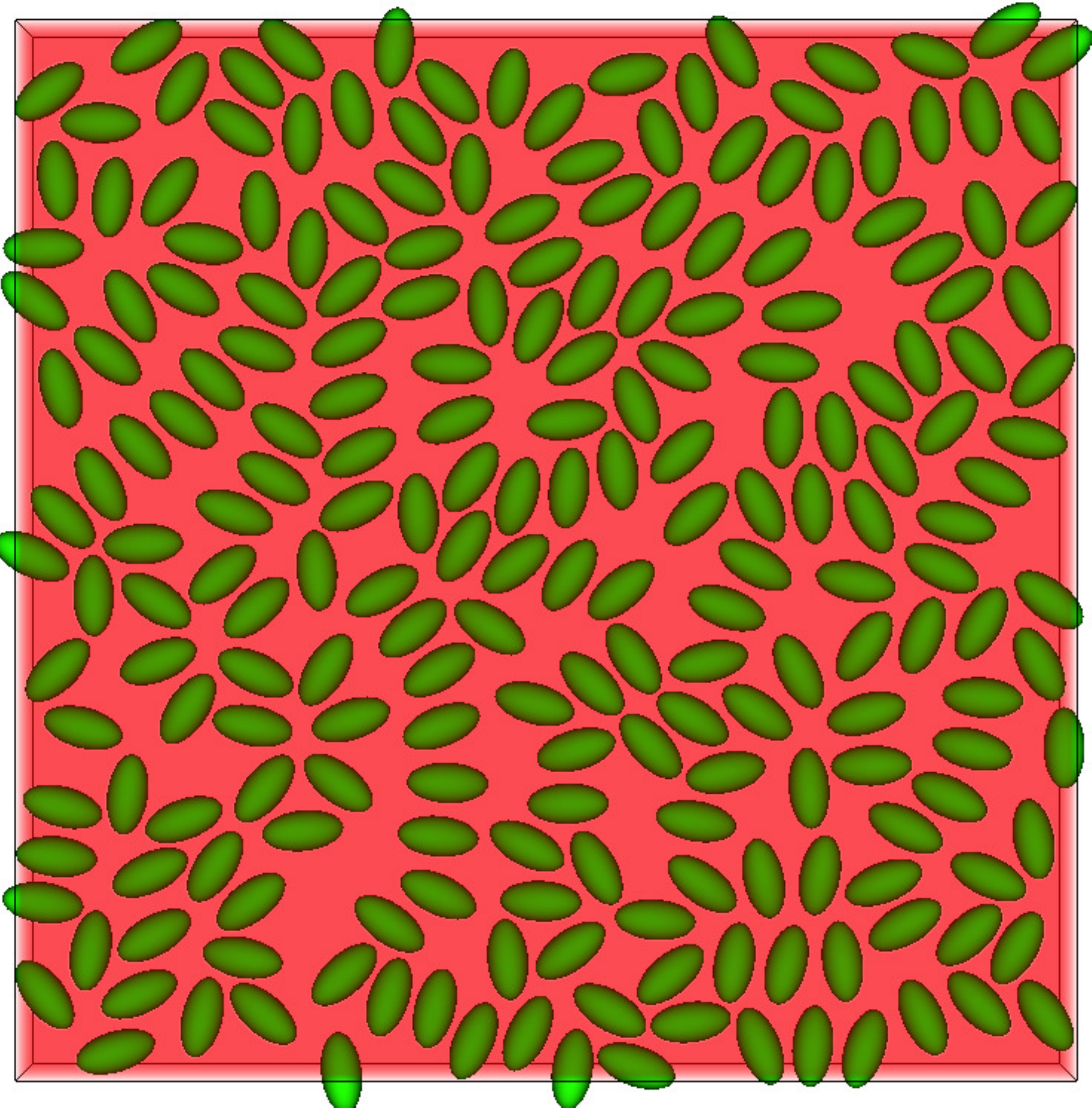}
\subcaption{} 
&\includegraphics[scale=0.135]{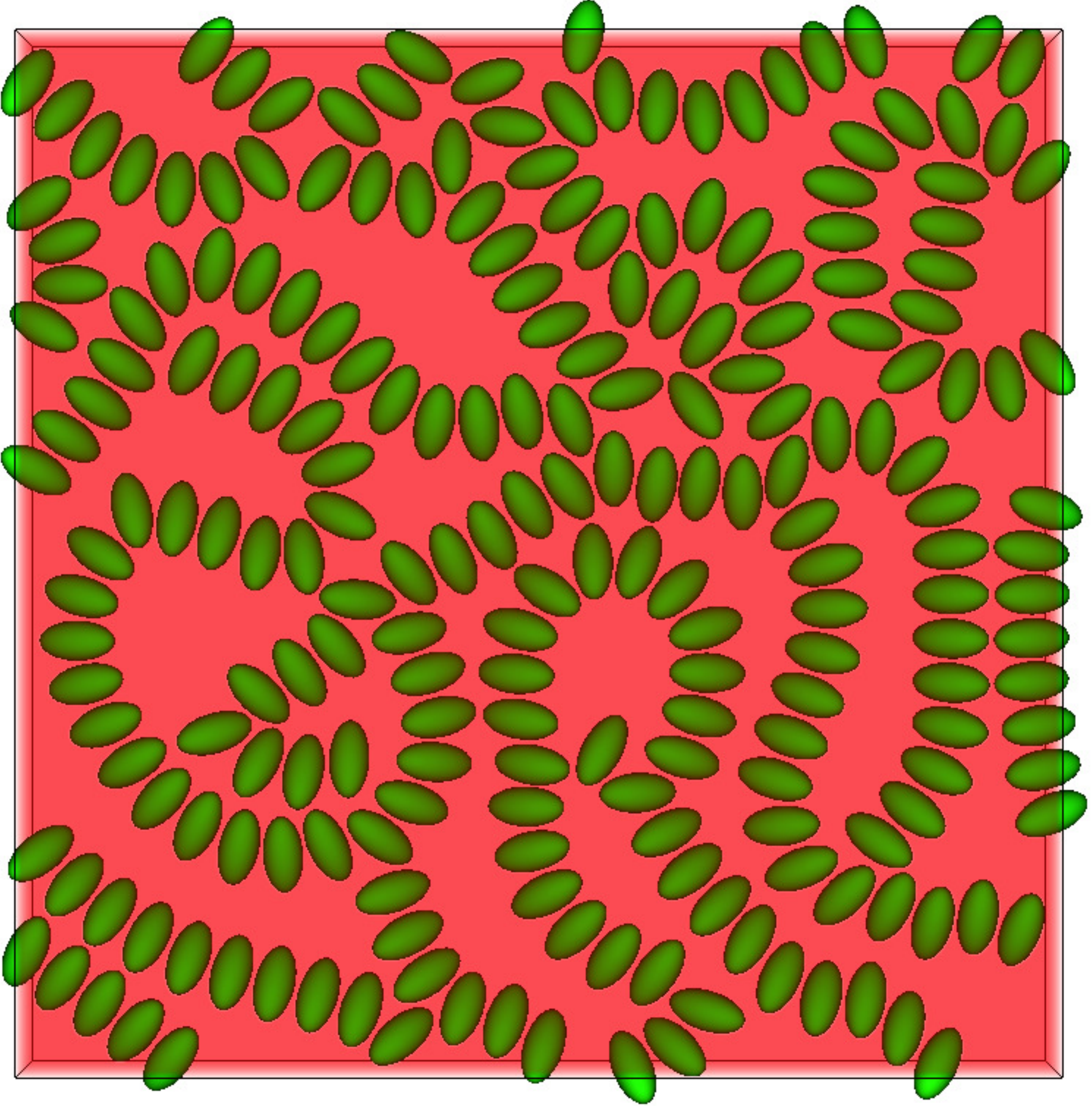}
\subcaption{} 
&\includegraphics[scale=0.135]{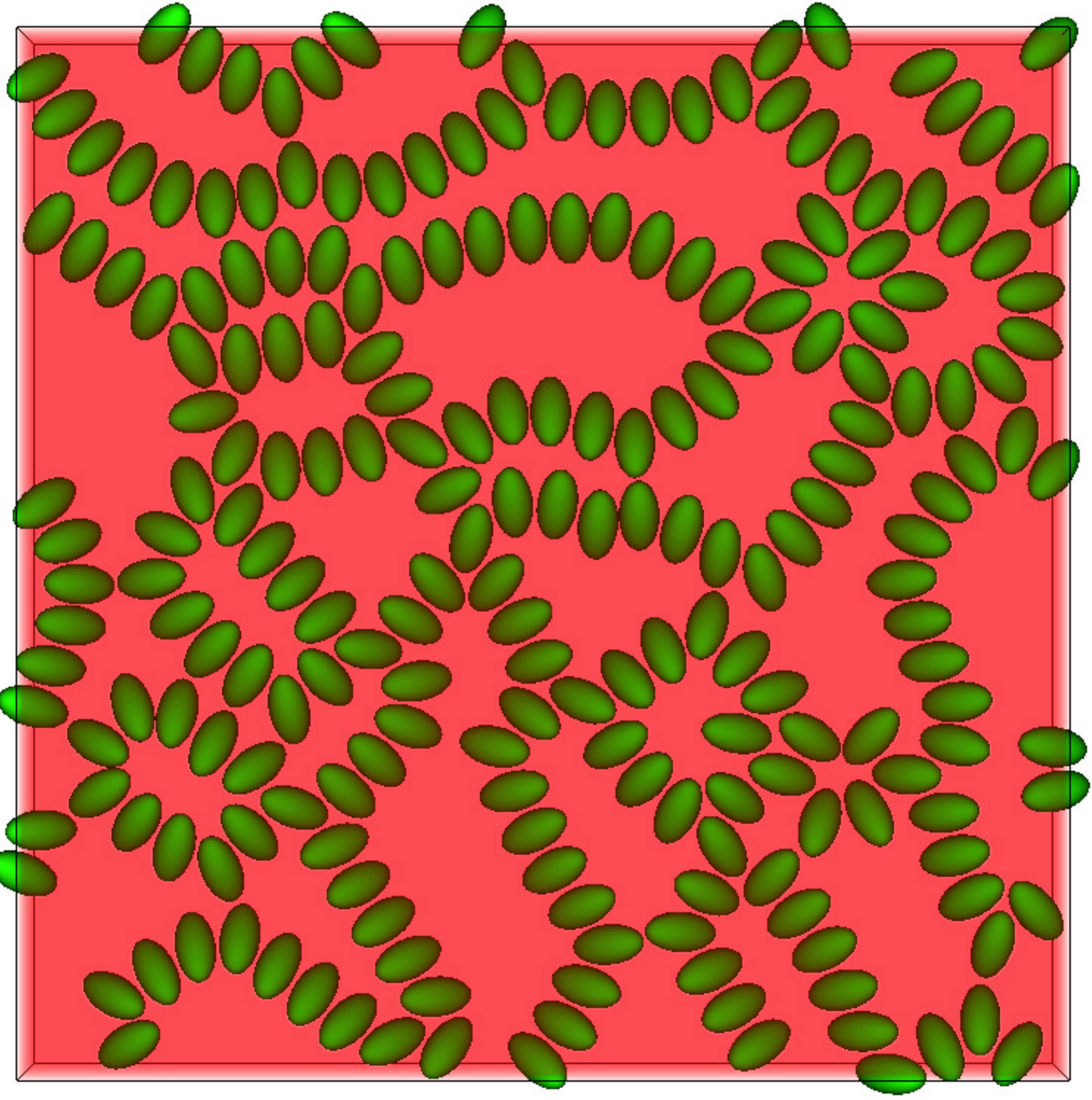}
\subcaption{} 
&\includegraphics[scale=0.135]{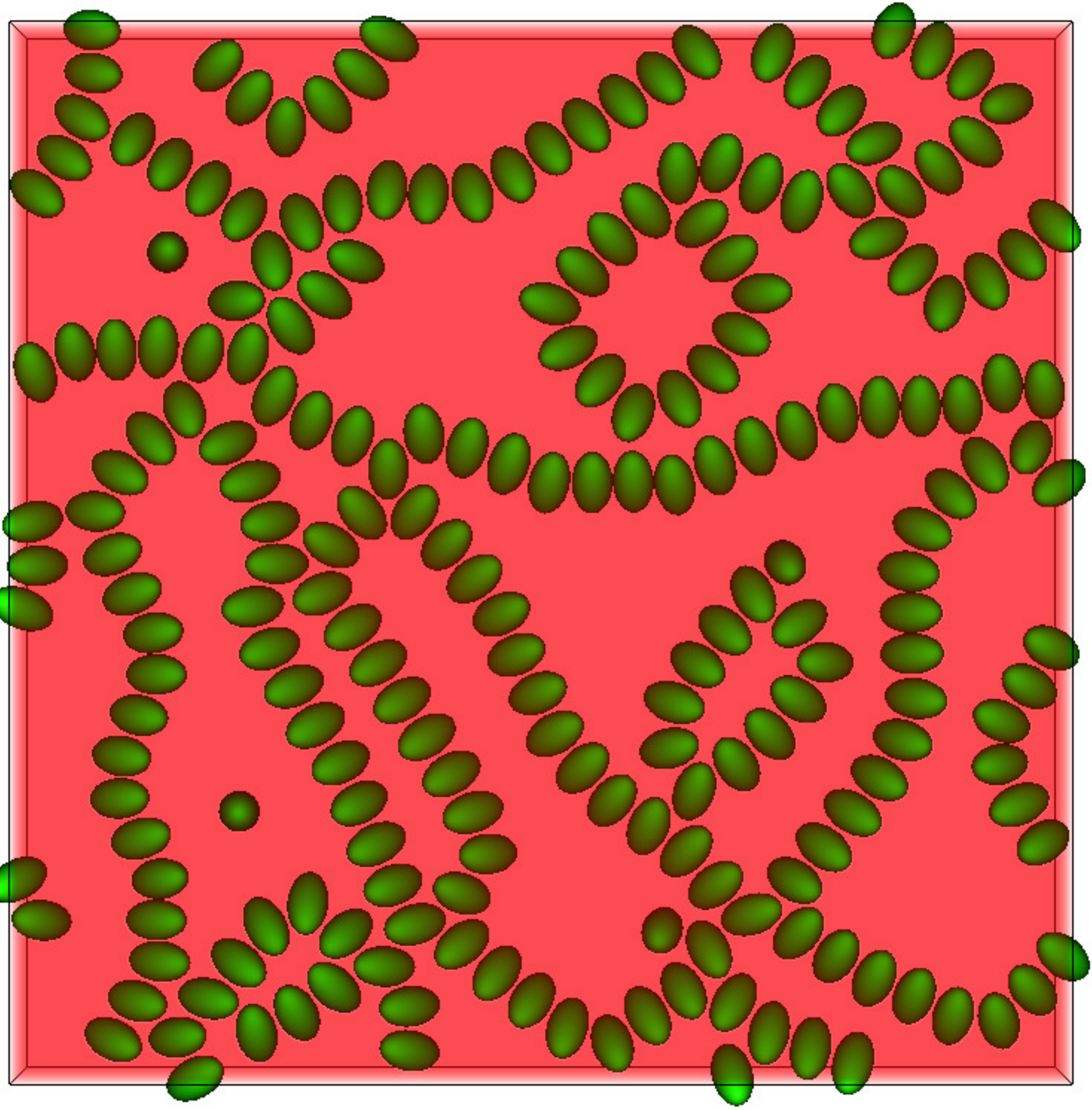}
\subcaption{} \\ [-0.9cm]
\rotatebox{90}{\hspace{1.5cm}$\phi=0.60$}
&\includegraphics[scale=0.135]{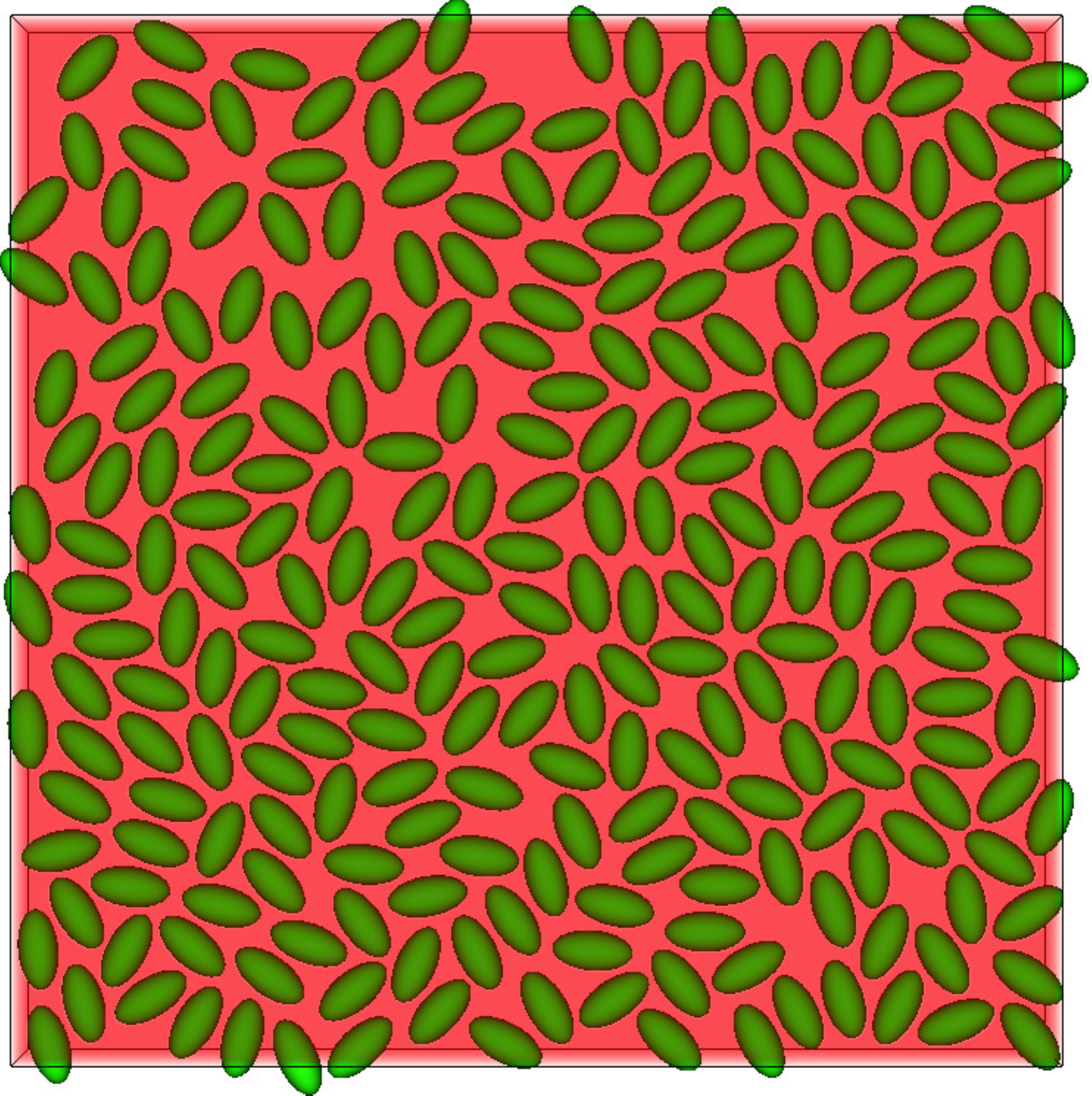}
\subcaption{} 
&\includegraphics[scale=0.135]{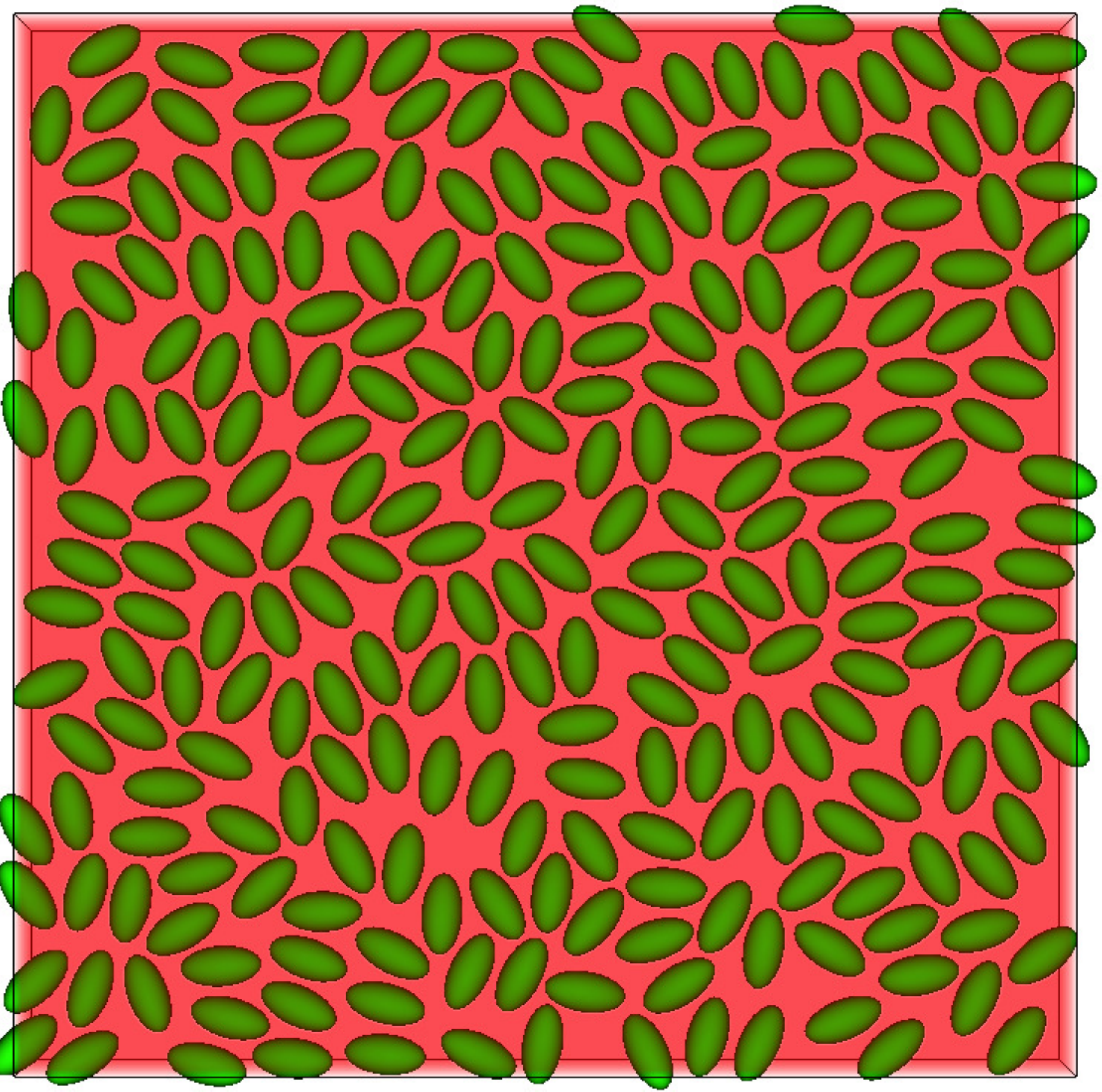}
\subcaption{} 
&\includegraphics[scale=0.135]{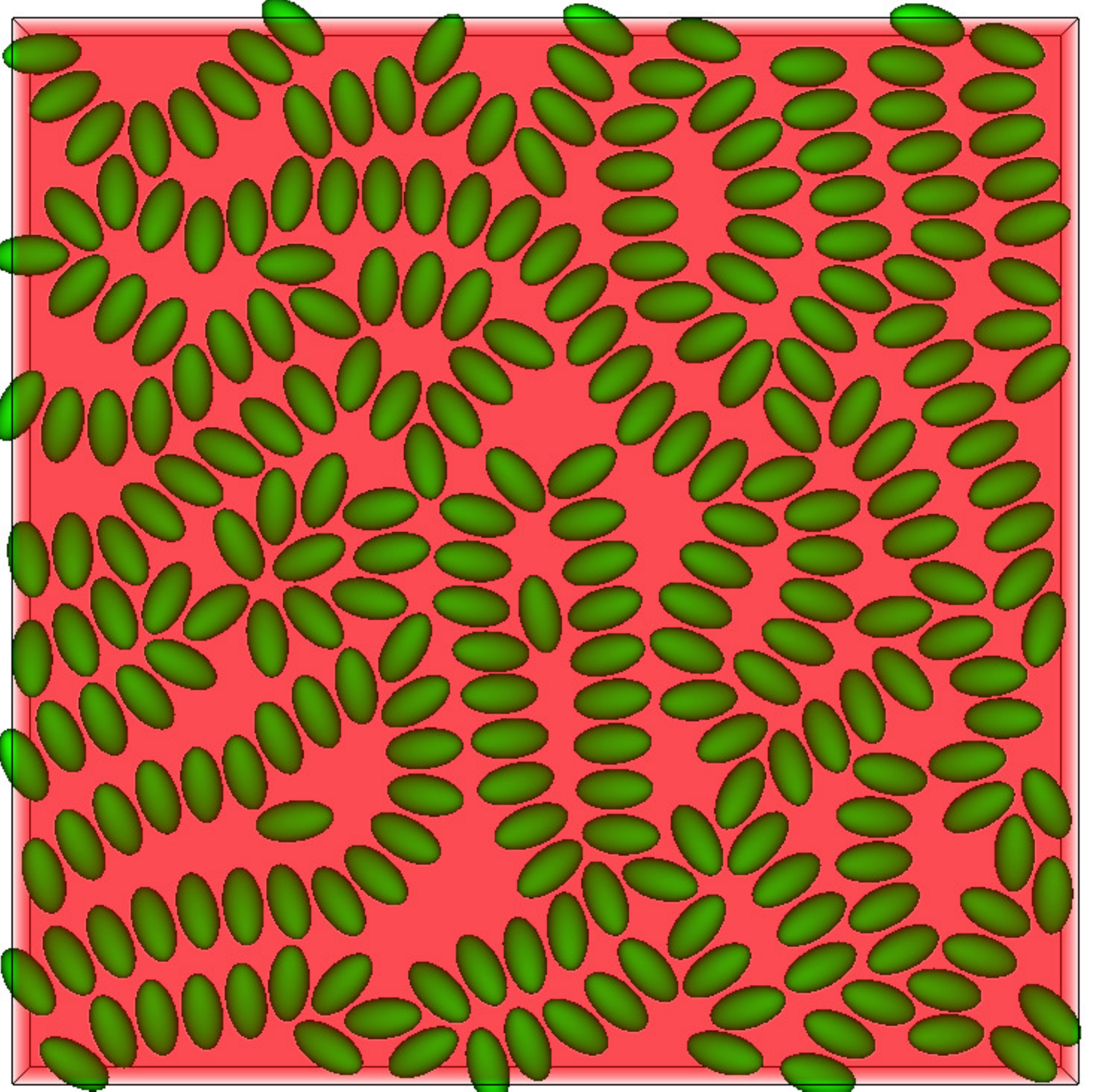}
\subcaption{} 
&\includegraphics[scale=0.135]{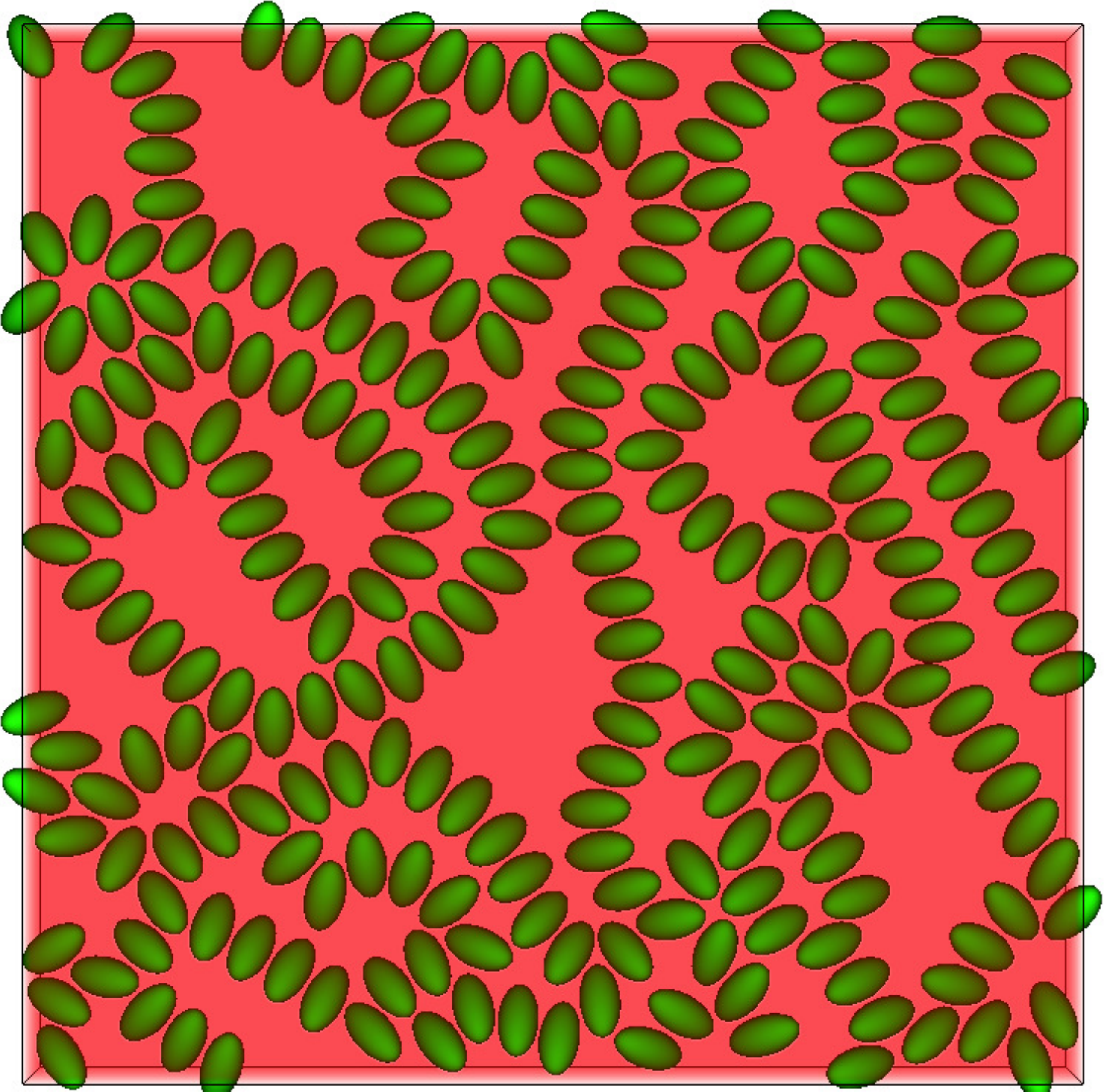}
\subcaption{} 
&\includegraphics[scale=0.135]{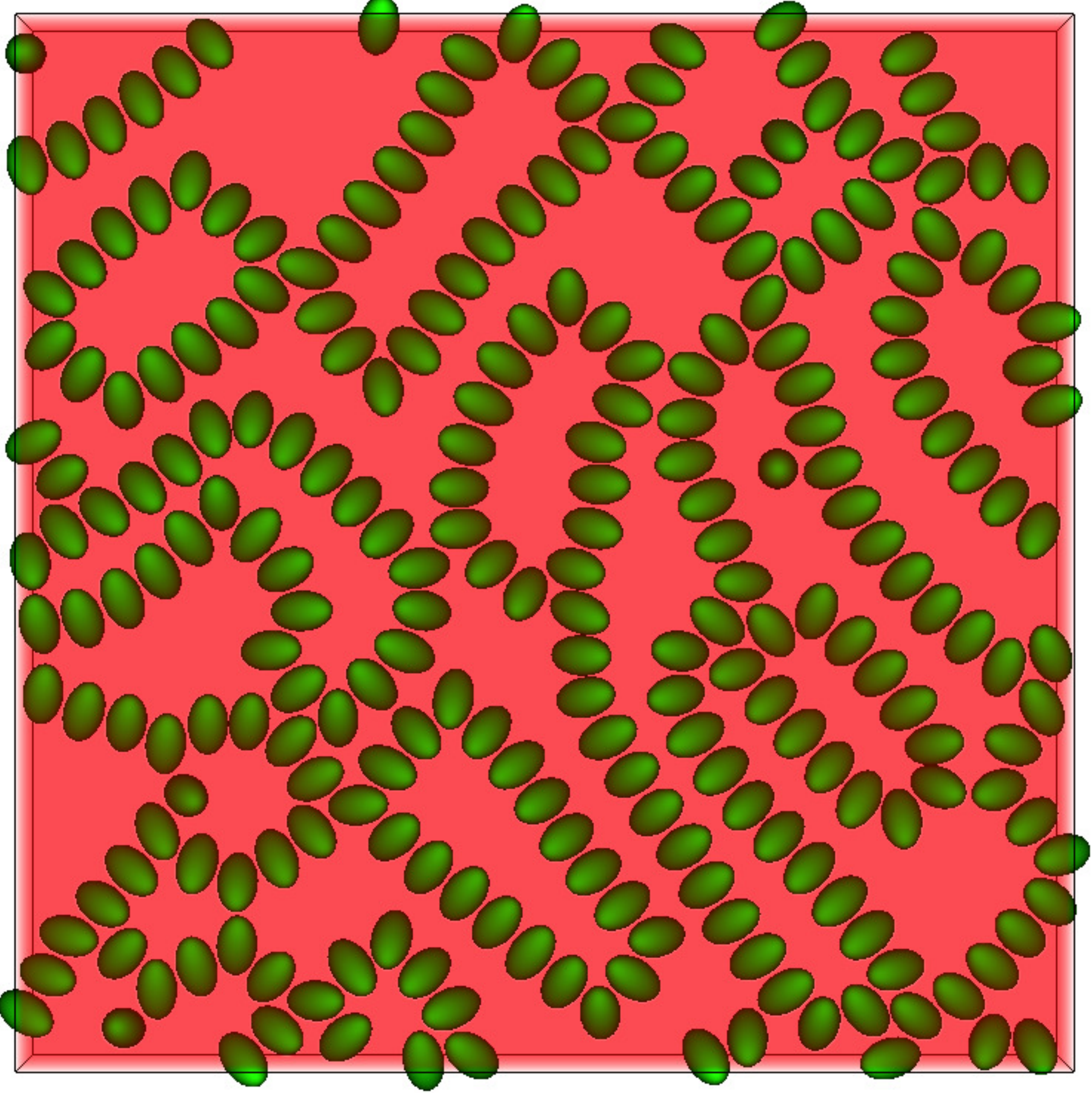}
\subcaption{}  \\ [-0.9cm]

\end{tabular}
\end{center}
\caption{\raggedright Snapshots of self-assembled structures as the external field, $B$, and surface fraction of particles, $\phi$, is varied, showing a strong dependence on both parameters. At small applied fields, $B=0.2B_c$, we see some particle ordering where particles prefer to be in a side-by-side or tip-to-tip state (b), (g), (l). For $B = 0.5B_c$ we observed the formation of curved ``capillary caterpillars'', in which the particles are oriented side-by-side and separate caterpillars prefer to face each other tip-to-tip (c), (h), (m). As the field strength is increased to $0.8B_c$, the caterpillars begin to prefer sharper, $90^{\circ}$ corners instead of curved chains (d), (i), (n). We see a further increase in corner sharpness for field strengths of $B = 1.2B_c$, and we observed small numbers of flipped particles in the vertical state for $\phi=0.38$ (e). As the surface fraction is increased to $\phi=0.53$ (j) and $\phi=0.60$ (o), fewer flipped particles are observed.}
\label{fig:panel}
\end{figure*}

\begin{figure}
	\includegraphics{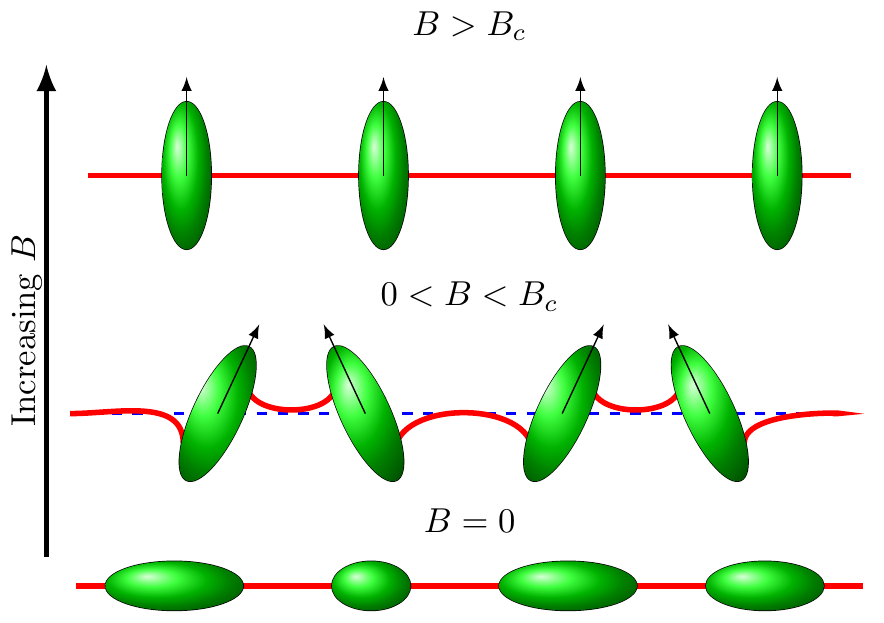}
	\caption{\raggedright Sideview illustrations. (Lower) The particles are randomly distributed on the interface, with $B=0$ (Figure~\ref{img:init}). (Middle) Due to capillary charges --- like menisci attract, unlike menisci repel --- the particles arrange into pairs called ``capillary couples'' because of the anti-symmetric nature of the meniscus formation (Figure~\ref{img:ordered1}). (Upper) Particles have transitioned to the vertical configuration, aligning with the external field. Interface deformations are absent due to the rotational symmetry of the three-phase contact-line and capillary interactions are turned off (Figure~\ref{img:ordered2} and~\ref{img:disordered}).}
	\label{img:sideview}
\end{figure}


When a magnetic prolate spheroidal particle with dipole moment, $\boldsymbol{\mu}$, is adsorbed at a fluid-fluid interface and subjected to an external magnetic field, $\mathbf{H}$, directed normal to the interface, it experiences a torque, $\boldsymbol{\mathrm{T}} = \boldsymbol{\mu} \times \mathbf{H}$, which attempts to align the particle with the external field. Surface-tension forces oppose the magnetic torque, and so for a given dipole-field strength, $B = \lvert \boldsymbol{\mu} \rvert \lvert \mathbf{H} \rvert$, the particle is tilted with respect to the external field, rather than aligned with it. When the particle is tilted with respect to the interface, the constant contact-angle condition stipulated by Young's equation means that the particle deforms the interface:~\cite{davies_effect_2014} the interface is depressed on one side and elevated on the other, as shown in Figure~\ref{img:single_particle}. 

When a critical dipole-field strength, $B_c$, is reached, however, the magnetic torque overcomes surface-tension forces and the particle undergoes a first-order phase transition and \textit{flips} from a tilted orientation to a vertical orientation, with respect to the interface. In the vertical orientation, interface deformations are absent because of the rotational symmetry of the three-phase contact-line in this configuration. 

The interface deformations that occur before the particle transitions to the vertical state are important because they are analogous to electrostatic charges, with a twist; depressions attract depressions, elevations attract elevations, and depressions repel elevations: opposites repel, rather than attract. Therefore, if more than one particle is adsorbed at a fluid-fluid interface and under the influence of an external magnetic field such that the dipole-field strength is less than the critical dipole-field strength, $B< B_c$, we expect these capillary charges to interact with each other.

Due to the anti-symmetric, dipolar nature of the interface deformations, these interactions are orientation dependent and will give rise to torques orthogonal to the interface causing in-plane rotation and ordering. Further, since the magnitude of the interface deformations depend on the dipole-field strength, it is possible to tune the strength of these dipolar capillary interactions by changing the strength of the external field, particle dipole moment, or both. 

We employed lattice-Boltzmann simulations~\cite{chen_lattice_1998, chen_recovery_1992,shan_lattice_1993,shan_simulation_1994,ladd_numerical_1994,ladd_numerical_1994-1,ladd_lattice-boltzmann_2001,bhatnagar_model_1954,orlandini_lattice_1995,swift_lattice_1996,jansen_bijels_2011,frijters_effects_2012,
davies_effect_2014} to investigate magnetic prolate spheroidal particles with aspect-ratio $\alpha=2$ adsorbed at a liquid-liquid interface under the influence of a magnetic field applied parallel to the interface normal, implemented in our LB3D code~\cite{lb3d_citation} with the same model and parameters as described in Davies et al.~\cite{davies_effect_2014} In our simulations, the dipole-field strength, $B$, is dominated by a strong external magnetic field and weak dipole moment so that magnetic dipole-dipole interactions are neglected. This means that the structures we observe are purely a result of dipolar capillary interactions between particles. 

In Figure~\ref{fig:self_assembly} we show the particles undergoing a first-order transition into the vertical orientation by exceeding the critical dipole-field strength for a single particle, $B_c$, demonstrating the unique phenomenology of the dipolar capillary interaction mode. The particles begin randomly oriented on the interface (Figure~\ref{img:init}). An external field is applied such that the dipole-field strength is $B = 0.5B_c$, creating capillary charges and causing the particles to interact with each other (Figure~\ref{img:ordered1}). The particles assemble into ``capillary caterpillars'', which we analyse further in Figure~\ref{fig:panel}. Finally, the dipole-field strength is increased beyond the single particle critical dipole field, $B > B_c$, where the particles transition into the vertical configuration (Figure~\ref{img:ordered2} and~\ref{img:disordered}) and capillary interactions are spontaneously switched off. 

After the particles have transitioned to the vertical orientation, they order according to a complex balance of other forces, such as magnetic dipole-dipole interactions, thermal fluctuations, and van der Waals forces, which depend strongly on different combinations of external field strength, dipole moment, particle size and shape, surface packing fraction and fluid properties, making final structures in the absence of capillary interactions difficult to predict. Figure~\ref{img:disordered} illustrates a situation in which thermal fluctuations are greater than magnetic dipole-dipole interactions, and the particles order randomly. \\
\indent The above mentioned parameters can be varied to observe the transition experimentally. For a typical system with a super-paramagnetic particle of long-axis radius $R_{\|} \approx 1\mathrm{\upmu m}$, saturated magnetic moment $\mu \approx 80 ~ \mathrm{A m^2 kg^{-1}}$ and surface-tension $\gamma \approx 0.05 ~ \mathrm{Nm^{-1}}$, a magnetic field strength $H < 1\mathrm{T}$ should suffice to observe the first-order phase transition.~\cite{bresme_orientational_2007} \\
\indent In Figure~\ref{fig:panel} we present a systematic investigation of the intermediate dipole-field strength regime before the particles have transitioned to the vertical configuration i.e. the particles are tilted with respect to the interface, for several surface-fractions, $\phi = N A_p/A$, where $N$ is the number of particles adsorbed at the interface, $A_p$ is the interface cross sectional area of a single particle, and $A$ is the area of the interface. \\
\indent We find that the self-assembled structures depend strongly on the dipole-field strength. For weak fields, $B=0.2B_c$, particles show some ordering, orienting in the side-by-side and tip-to-tip configuration (Figure~\ref{fig:panel}b,~\ref{fig:panel}g, and~\ref{fig:panel}l). As the dipole-field strength is increased to $B=0.5B_c$ the particles begin to form long, curved ordered chains, or ``capillary caterpillars'', in which particles strongly prefer to orient side-by-side (Figure~\ref{fig:panel}c,~\ref{fig:panel}h, and~\ref{fig:panel}m). Different capillary caterpillars face each other in a tip-tip configuration, as can be seen in the upper-right of Figure~\ref{fig:panel}m. The particles in this region align in ``capillary couples''  (Figure~\ref{img:sideview}). 

For dipole-field strengths of $B=1.2B_c$, the particles should be flipped into the vertical state having exceeded the critical dipole-field strength required to transition a single particle into the vertical state (Figure~\ref{img:sideview} and~\ref{img:ordered2}). However, for low surface fractions $\phi=0.38$ only a small number of particles are flipped (Figure~\ref{fig:panel}e), and the number of particles in the flipped state decreases as the surface-fraction increases to $\phi=0.53$ (Figure~\ref{fig:panel}j) and $\phi=0.60$ (Figure~\ref{fig:panel}o). This suggests that many-body effects inhibit the first-order phase transition and shift the critical dipole-field required to flip the particles into the vertical state to larger values.

Additionally, for dipole-field strengths of $B=0.8B_c$ (Figure~\ref{fig:panel}d,~\ref{fig:panel}i, and~\ref{fig:panel}n) and $B=1.2B_c$ (Figure~\ref{fig:panel}e,~\ref{fig:panel}j, and~\ref{fig:panel}o) the capillary caterpillars favour straight chains with sharp corners rather than curved chains as seen for $B=0.5B_c$ (Figure~\ref{fig:panel}b,~\ref{fig:panel}g, and~\ref{fig:panel}l). 
A possible explanation is as follows. For a larger dipole-field strength, the particles' tilt-angle with respect to the interface increases.~\cite{davies_effect_2014} Particles prefer to align with their dipole axes parallel with one another. The sharper corners are simply changes to the in-plane components of the dipole-dipole angle due to the larger tilt-angles, however, a more thorough investigation of the formation and properties of individual capillary caterpillars is needed to confirm this hypothesis.\\\indent
Compared with the self-assembled structures observed due to quadrupolar capillary interactions, \cite{loudet_capillary_2005,loudet_self-assembled_2009,loudet_how_2011,botto_capillary_2012,botto_capillary_rod_2012} we find that the dipolar capillary interactions also favour side-side orientations. Only once individual capillary caterpillars have formed do the particles in each caterpillar arrange tip-tip with particles in other caterpillars. Fully understanding dipolar capillary-induced chain formation is a priority for future study. For quadrupolar capillary interactions, tip-tip configurations occur when electrostatic repulsion between particles exists, and capillary arrows can form when the particles are of unequal size.~\cite{loudet_capillary_2005,loudet_self-assembled_2009,loudet_how_2011} The effect of particle size, aspect-ratio, contact-angle, charge and magnetic moment remain unexplored for the dipolar capillary interactions we report here.  \\\indent
By using magnetic anisotropic particles interacting with external magnetic fields, we showed how to dynamically tune dipolar capillary interactions between particles by varying the dipole-field strength, and how to switch these dipolar capillary interactions on and off by making the particles undergo a first-order orientation transition. Our simulations reveal novel self-assembled structures that depend on the surface coverage of particles and the dipole-field strength. We observed the formation of ``capillary caterpillars'', in which particles align in side-side configurations, and ``capillary couples'' where particles in individual caterpillars align in tip-tip chains with particles in other caterpillars, due to the anti-symmetric menisci formation. In addition to providing motivation for new experiments on the bottom-up fabrication of new materials, the novel assembly behaviour reported here should also be observable in colloid-liquid crystal mixtures,~\cite{meeker_colloidliquid-crystal_2000,cleaver_network_2004} and has implications for liquid crystals in general. Further, the discontinuous transition of the prolate spheroidal particles and the on-off tunability of capillary interactions could find applications in photonic systems that require dynamic control of optical properties, such as e-readers.~\cite{kim_self-assembled_2011} Finally, the sensitivity of the first order orientation transition, and hence the assembly process, to the particle size and shape, external field strength and interface surface tensions has potential applications in colloidal metrology for sensors that respond to small changes in interface properties. 

\begin{acknowledgments}
GBD \& PVC are grateful to EPRSC and Fujitsu Laboratories Europe for funding GBD's PhD studentship. JH acknowledges financial support from NWO/STW (VIDI grant 10787 of J. Harting). FB thanks EPSRC for awarding a Leadership Fellowship and TK acknowledges the University of Edinburgh for awarding a Chancellor's Fellowship.
We ran our simulations on HECToR and ARCHER, the UK’s HPC service, via CPU time awarded by grants EPSRC ``Large Scale Lattice Boltzmann for Biocolloidal Systems” (EP/I034602/1), ``2020 Science" (EP/I017909/1), ``UK Consortium on Mesoscale Engineering Sciences"  (EP/L00030X/1) and EU-FP7 CRESTA grant (Grant No. 287703).
\end{acknowledgments}

\bibliography{torquebib,tnopapers}

\end{document}